# Topological Type-II Dirac Fermions Approaching the Fermi Level in a Transition Metal Dichalcogenide NiTe$_2$

Chunqiang Xu[†,‡], Bin Li[§,‡], Wenhe Jiao[¥], Wei Zhou[†], Bin Qian[†], Raman Sankar[$,⊥,*], Nikolai D. Zhigadlo[£], Yanpeng Qi[#], Dong Qian[&], Fang-Cheng Chou[$], Xiaofeng Xu[†,*]

[†]Advanced Functional Materials Lab, Department of Physics, Changshu Institute of Technology, Changshu 215500, China

[§]Information Physics Research Center, Nanjing University of Posts and Telecommunications, Nanjing, 210023, China

[¥]Department of Physics, Zhejiang University of Science and Technology, Hangzhou 310023, China

[$]Center for Condensed Matter Sciences, National Taiwan University, Taipei 10617, Taiwan

[⊥]Institute of Physics, Academia Sinica, Nankang, Taipei R.O.C. Taiwan 11529

[£]Department of Chemistry and Biochemistry, University of Berne, Freiestrasse 3, CH-3012 Berne, Switzerland

[#]School of Physical Science and Technology, ShanghaiTech University, Shanghai, 201210，China.

[&]Key Laboratory of Artificial Structures and Quantum Control (Ministry of Education), School of Physics and Astronomy, Shanghai Jiao Tong University, Shanghai 200240, China

**ABSTRACT:** Type-II Dirac/Weyl semimetals are characterized by strongly tilted Dirac cones such that the Dirac/Weyl node emerges at the boundary of electron and hole pockets as a new state of quantum matter, distinct from the standard Dirac/Weyl points with a point-like Fermi surface which are referred to as type-I nodes. The type-II Dirac fermions were recently predicted by theory and have since been confirmed in experiments in the PtSe$_2$-class of transition metal dichalcogenides. However, the Dirac nodes observed in PtSe$_2$, PdTe$_2$ and PtTe$_2$ candidates are quite far away from the Fermi level, making the signature of topological fermions obscure as the physical properties are still dominated by the non-Dirac quasiparticles. Here we report the synthesis of a new type-II Dirac semimetal NiTe$_2$ in which a pair of type-II Dirac nodes are located very close to the Fermi level. The quantum oscillations in this material reveal a nontrivial Berry's phase associated with these Dirac fermions. Our first principles calculations further unveil a topological Dirac cone in its surface states. Therefore, NiTe$_2$ may not only represent an improved system to formulate the theoretical understanding of the exotic consequences of type-II Dirac fermions, it also facilitates possible applications based on these topological carriers.

## 1. INTRODUCTION

Following the discovery of topological insulators, i.e., an insulating bulk enclosed by topologically protected metallic boundaries, the search for quantum materials with diverse symmetry-protected topological phenomena has become one of the central topics in condensed matter physics.[1,2] Soon after, the advent of 3D Dirac and Weyl semimetals, now often referred to as type-I fermions in which the Dirac/Weyl node is enclosed by an electron/hole pocket, has triggered a tremendous research interest due to the rich physics and the potential applications therein.[3] Very recently, the focus of the field has partially shifted to the so-called type-II Dirac/Weyl systems, where the Dirac cone is strongly tilted as a result of broken Lorentz symmetry, and as such the Dirac/Weyl point appears only at the contact of electron and hole pockets.[4] The different band topology in type-II Dirac/Weyl semimetals can lead to distinct physical properties, such as the direction-dependent chiral anomaly,[5,6] exotic superconductivity[7,8] and novel quantum oscillations.[9]

The material incarnations of these type-II Dirac/Weyl fermions include the transition metal dichalcogenides like (W, Mo)Te$_2$,[10,11] the diphosphides (Mo, W)P$_2$,[12-14] Ta$_3$S$_2$[15] and LaAlGe[16] etc. More recently, the transition metal dichalcogenides (TMDs) TMX$_2$ (TM=Pd, Pt; X=Se, Te) were proposed to host type-II Dirac fermions and were soon verified in angle-resolved photoemission spectroscopy (ARPES) experiments.[17-28] Interestingly, PdTe$_2$ also superconducts below $T_c$=2 K and the Au substitution can boost $T_c$ to a maximum value of 4.7 K, making this family of TMDs a promising platform to search for topological superconductors.[29-34] However, the Dirac points reported in these materials are located deeply below the Fermi level, with a binding energy of 0.6 eV in PdTe$_2$, 0.8 eV in PtTe$_2$, and 1.2 eV in PtSe$_2$.[20,22,24] The Dirac node far from the Fermi level smears out the signature of relativistic particles and makes the physical properties mainly dominated by the nonrelativistic carriers, evidenced by, as an example, the quadratic increase of the magnetoresistance (MR) rather than the linear MR often seen in Dirac systems in the quantum limit.[35-39]

Surprisingly, the Ni-based counterpart NiTe$_2$ has not been studied so far, in the context of topological physics. This material was synthesized about six decades ago and the previous reports only presented its crystal structure [40-45] or gave the position and character of the metal $d$ bands at the Γ point.[46,47] Its band features, in particular the Dirac dispersion if any, have not been investigated from a topological perspective. In this paper, we report the synthesis and the characterization of high quality single crystals of NiTe$_2$. The transport of this material shows the evidence of non-saturating linear MR, characteristic of topological metals. The quantum oscillations reveal a light effective mass and a nontrivial Berry phase for the associated carriers. Like other PdTe$_2$-class of TMDs, the electronic structure calculations also manifest a tilted band crossing, confirming the type-II nature of the Dirac fermions. Compared to other compounds in this series, the Dirac node in NiTe$_2$ is much closer to the Fermi level, which may be responsible for the more prominent signatures of topological carriers in this material. Our calculations further uncover the existence of topologically nontrivial surface states in this compound.[19,48]

## 2. EXPERIMENT

The single crystals of NiTe$_2$ were synthesized in Te solution. Accurately weighed amounts of high purity nickel powder and tellurium ingots were mixed thoroughly with a molar ratio of 1:8 in a glove box and then sealed in an evacuated quartz tube. This quartz tube was then heated to 950 °C quickly in a sintering furnace and kept at this temperature for 48 h, before being slowly cooled down to 500 °C (3 °C /h), and finally being quenched into cold water. The excess amount of Te was centrifuged at 500 °C. The air stable, large pieces of dark-gray layered single crystals of NiTe$_2$ of typical 7-8 mm in length were harvested.

X-ray diffraction (XRD) measurements were performed at room temperature using a Rigaku diffractometer with Cu $K\alpha$ radiation and a graphite monochromator. Lattice parameters were obtained by Rietveld refinements. In order to check the actual chemical composition of the samples, Energy Dispersive X-ray (EDX) Spectroscopy analysis was performed on a set of crystals. The chemical off-stoichiometry was seen to be very small (within 1%) from the multiple-spot scanning on the surface with an incident beam energy of 20 keV. For each sample studied, we chose ~20 spots of different sizes on the sample surfaces to do the EDX analysis thus confirmed the stoichiometry of the samples. Raman spectroscopy was also performed at room temperature. The resistivity and the magneto-transport measurements were carried out in a Physical Property Measurement System (PPMS-9, Quantum Design) with $ac$ transport option (typical excitation current I=1 mA). The standard 4-probe method was employed for the resistivity measurements. The Hall effect was performed by reversing the field direction and antisymmetrizing the data. The specific heat data were also collected on PPMS. For the thermoelectric power (TEP) and the Nernst measurements, a modified steady-state method was used in which a temperature gradient, measured using a constantan-chromel differential thermocouple, was set up across the sample via a chip heater attached to one end of the sample. The thermopower/Nernst voltage was read out by a nanovoltmeter K2182A from Keithley Instruments. The magnetization data were measured using Quantum Design MPMS (7 T) system. Torque magnetometry measurements were carried out on a piezoresistive cantilever which was mounted on a rotation probe. The samples were mounted onto the cantilever using nonmagnetic grease.

We computed the electronic structures with high accuracy using the full-potential linearized augmented plane wave (FP-LAPW) method implemented in the WIEN2K code.[49] The generalized gradient approximation (GGA) was applied to the exchange-correlation potential calculation.[50] The muffin-tin radii were chosen to be 2.31 a.u. for both Ni and Te. The plane-wave cutoff was defined by $RK_{max}$ = 7.0, where $R$ is the minimum LAPW sphere radius and $K_{max}$ is the plane-wave vector cutoff. Spin-orbit coupling was included in the calculations. To calculate the surface electronic structure and Fermi surface map, we constructed a first-principles tight-binding model Hamilton, where the tight-binding model matrix elements were calculated by projecting onto the Wannier orbitals.[51] We used Ni $s$ and $d$, Te $s$ and $p$ orbitals to construct Wannier functions. The surface state spectrum of (001) slab were calculated with the surface Green's function methods as implemented in WannierTools.[52] Fermi surfaces were visualized with the XCrySDen program.[53] The de Haas-van Alphen (dHvA) frequencies and band masses were calculated using the SKEAF (Supercell K-space Extremal Area Finder) algorithm at an interpolation of 150 in the full Brillouin zone (BZ).[54]

## 3. RESULTS AND DISCUSSION

NiTe2 crystallizes in the CdI$_2$-type trigonal structure with $P\bar{3}m1$ space group, as schematically shown in Figure 1a. The atomic structure consists of NiTe$_2$ layers stacked along the $c$-axis. Each layer is composed of the edge-sharing NiTe$_6$ octahedra with Ni octahedrally coordinated by 6 Te atoms. The EDX spectrum (Figure 1b) confirms the chemical stoichiometry of the as-grown samples. The Ramam shift in the inset shows a peak around 76 cm$^{-1}$ which is probably from the $E_g$ vibrational mode for 1T structure. The sharp XRD peaks (Figure 1c) indicate the high quality of the samples. The powder XRD patterns in Figure 1d show no secondary phase and can be well indexed. The Rietveld refinements give the lattice parameters of $a$ = 3.8550(9) Å, $c$ = 5.2659(8) Å, consistent with the stoichiometric NiTe$_2$ in the literature[44,45]. More details on the XRD refinement can be found in the Supplementary Information.

The temperature evolution of in-plane resistivity for NiTe$_2$ under various fields is illustrated in Figure 2a. At zero field, the sample is metallic down to the lowest temperature studied (2 K), with the residual resistivity ratio (RRR) an order of 400, and no superconductivity is observed in this $T$ regime. With increasing field above 5 T, however, the $\rho(T)$ curves pass through a minimum and show pronounced upturns at low temperatures, as often seen in many topological materials in the magnetic field. The heat capacity characterization of the sample shows no evident anomaly below 200 K (Figure 2b). The separation of electronic ($\gamma T$) and phononic ($\beta T^3$) contributions at low temperatures yields $\gamma$=5.71 mJ/mol K$^2$ and $\beta$=0.64 mJ/mol K$^4$ (the Debye temperature $\Theta_D$=230 K). The MR and the Hall resistivity $\rho_{xy}$ are depicted in panel **c** and **d**, respectively. As noticed, the MR of this paramagnetic material is rather large[40], reaching as high as 1250% at 2 K and 9 T. Importantly, the MR at low $T$ is quite linear, a feature seen in many topological materials due to the distinct spectrum of Landau levels for Dirac fermions in the field.[35,36,38] A minor quadratic contribution in the MR can also be observed, especially at higher temperatures. For standard metals with both electrons and holes, the MR increases quadratically at low field and tends to saturate at high field, i.e., MR scales as $\frac{\alpha B^2}{\beta + B^2}$.[55,56] Hence the overall MR data with both Dirac fermions and conventional carriers can be fitted by a dominant $B$–linear term in addi-tion to a small conventional component, i.e., $\frac{\Delta\rho}{\rho} = \frac{\alpha B^2}{\beta + B^2} + \gamma B$. As seen from Figure 2c, this dichotomy between Dirac electrons and conventional carriers fits the experimental data very well. The Hall resistivity, on the other hand, can be modelled with a two-band carrier model, as seen from the fit in Figure 2d. As noted, no Shubnikov-de Hass (SdH) oscillations are observed either in the MR or in the Hall resistivity up to 9 T. However, dHvA oscillations, to be discussed in detail below, appear in a field as low as 3 T. This inconsistency between SdH and dHvA oscillations is often seen in low-dimensional materials and arises from the distinct mechanisms of SdH and dHvA oscillations. While SdH oscillations come from the oscillating scattering rate and can thus be complicated by the detailed scattering processes, the dHvA effect is caused directly by the free energy oscillations of a system. In Figure 2e, the TEP is positive in the whole $T$-range measured, consistent with the Hall coefficient. The TEP is characterized by a large peak around 25 K, presumably due to the phonon-drag effect.[57] A 4 T magnetic field is seen to have little bearing on the TEP (Figure 2e). As seen in Figure 2f, the Nernst signal ($S_{xy} = \frac{E_y}{\nabla T}$) is linear in field and the slope corresponds to the Nernst coefficient ($\nu = \frac{S_{xy}}{B}$) given in the inset of Figure 2f. It is evident that the Nernst coefficient increases remarkably below ~50 K, possibly because of the significant rise of carrier mobility at low temperatures.[58,59]

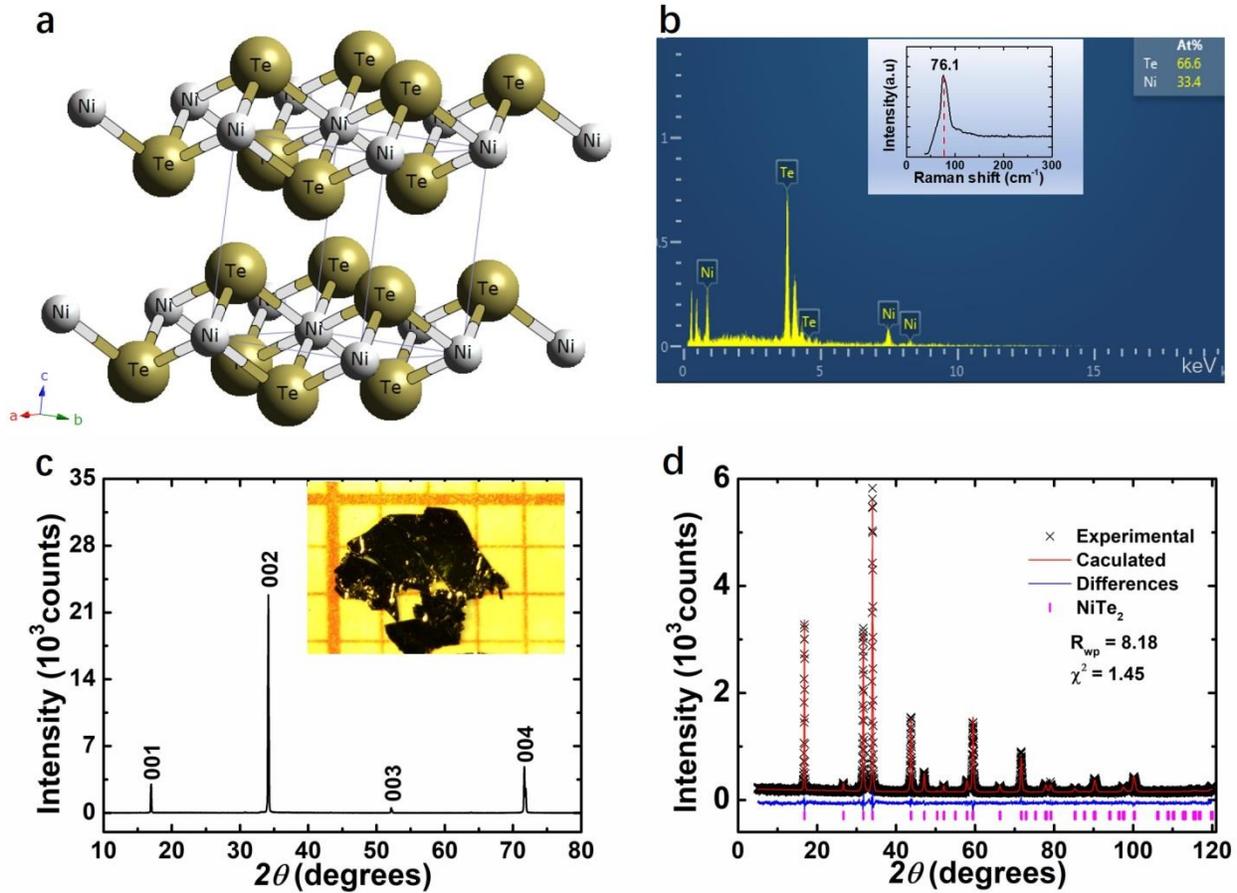

**Figure 1.** Structural characterization of NiTe2. **a** The crystal structure of NiTe$_2$. **b** The EDX of the sample. The inset gives the Raman spectroscopy. **c** The XRD pattern for a single crystal. The inset is the optic image of the as-grown crystals. **d** The powder XRD patterns and the Rietveld refinements. The refinement results are given in the Supplementary Information.

The $T$-dependent magnetization, measured under a field of 1 Tesla, is shown in Fig. 3a. In the whole temperature range studied, the sample is paramagnetic. Below $T$ ~25 K, the magnetization undergoes a rapid increase. The plot of reciprocal magnetization (right axis of Fig. 3a) reveals the Curie-Weiss law between ~150 K to ~25 K (the red dotted line). The negative Curie temperature -134 K as extracted indicates the antiferromagnetic interaction of the local moments that is responsible for the magnetization upturn at low temperatures,[40] although no magnetic order is observed down to 2 K. The isothermal magnetization for $H \parallel ab$ at several temperatures shows beautiful quantum oscillations, which can directly measure the topology of the underlying Fermi surfaces. The oscillatory components after subtracting the background are shown in Figure 3b. The Fast Fourier transform (FFT) of the dHvA oscillations is shown in Figure 3c. Only two major bands are observable in the range up to 500 T, labeled hereafter as $\alpha$ and $\beta$ corresponding to the frequencies ($F$) of 50 T and 394 T, respectively. In general, the oscillatory dHvA data can be described by the standard Lifshitz-Kosevich (LK) formula,[60]

$$\Delta M \propto -R_T \cdot R_D \cdot R_S \cdot \sin\left[2\pi\left(\frac{F}{B} + \frac{1}{2} - \frac{\phi_B}{2\pi} - \delta\right)\right] \quad (1)$$

where $R_T$, $R_D$, $R_S$ are the thermal damping factor, Dingle damping term and a spin-related damping term, respectively. $R_T = \frac{2\pi^2 k_B T m^*/eB\hbar}{\sinh(2\pi^2 k_B T m^*/eB\hbar)}$, $R_D = exp(2\pi^2 k_B T_D m^*/eB\hbar)$, and $R_s = \cos(\pi g m^*/2 m_e)$, where $m^*$ is the effective electron mass, $m_e$ is free electron mass, $k_B$ is the Boltzmann constant and $T_D$ is the Dingle temperature. $\phi_B$

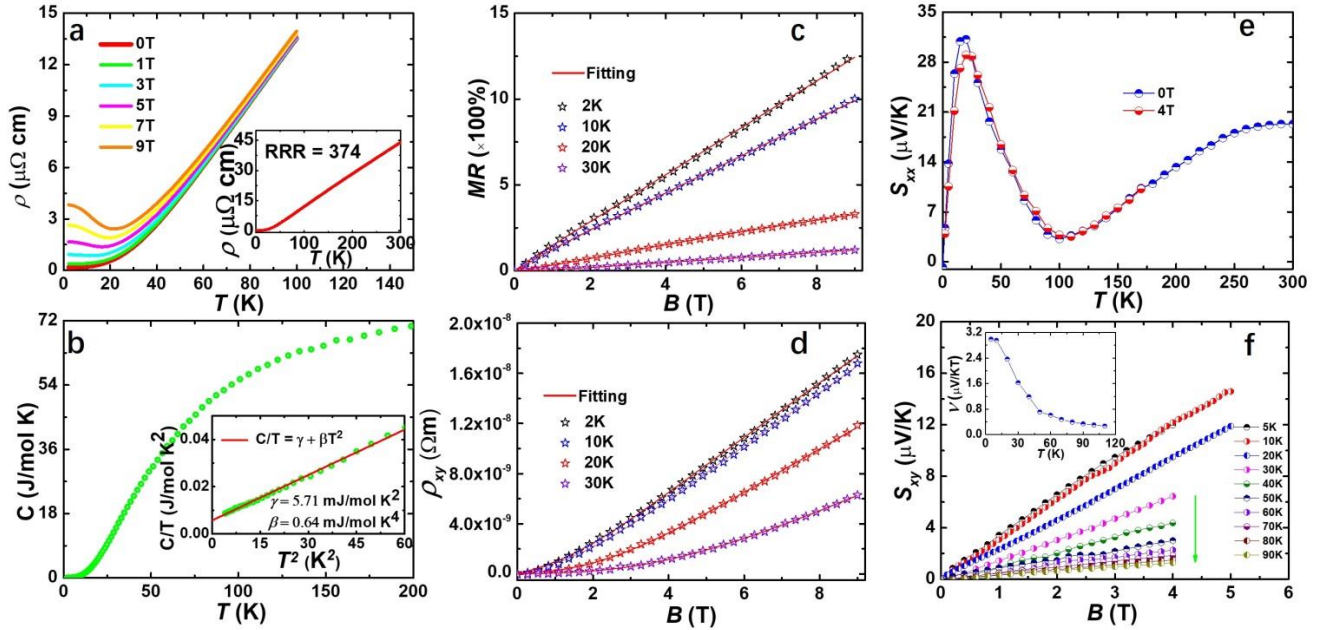

**Figure 2.** Physical properties of NiTe2. **a** The resistivity at some selected fields. The inset plots the zero-field resistivity separately. **b** The heat capacity at zero field below 200 K. The inset separates the electronic and phononic contributions. **c** and **d** The MR

and the Hall resistivity at some selected temperatures. The red solid lines are the fits (see main text). **e** The TEP at both 0 T and 4 T. **f** The Nernst signals as a function of field at different temperatures. The inset gives the Nernst coefficient ($v = \frac{S_{xy}}{B}$).

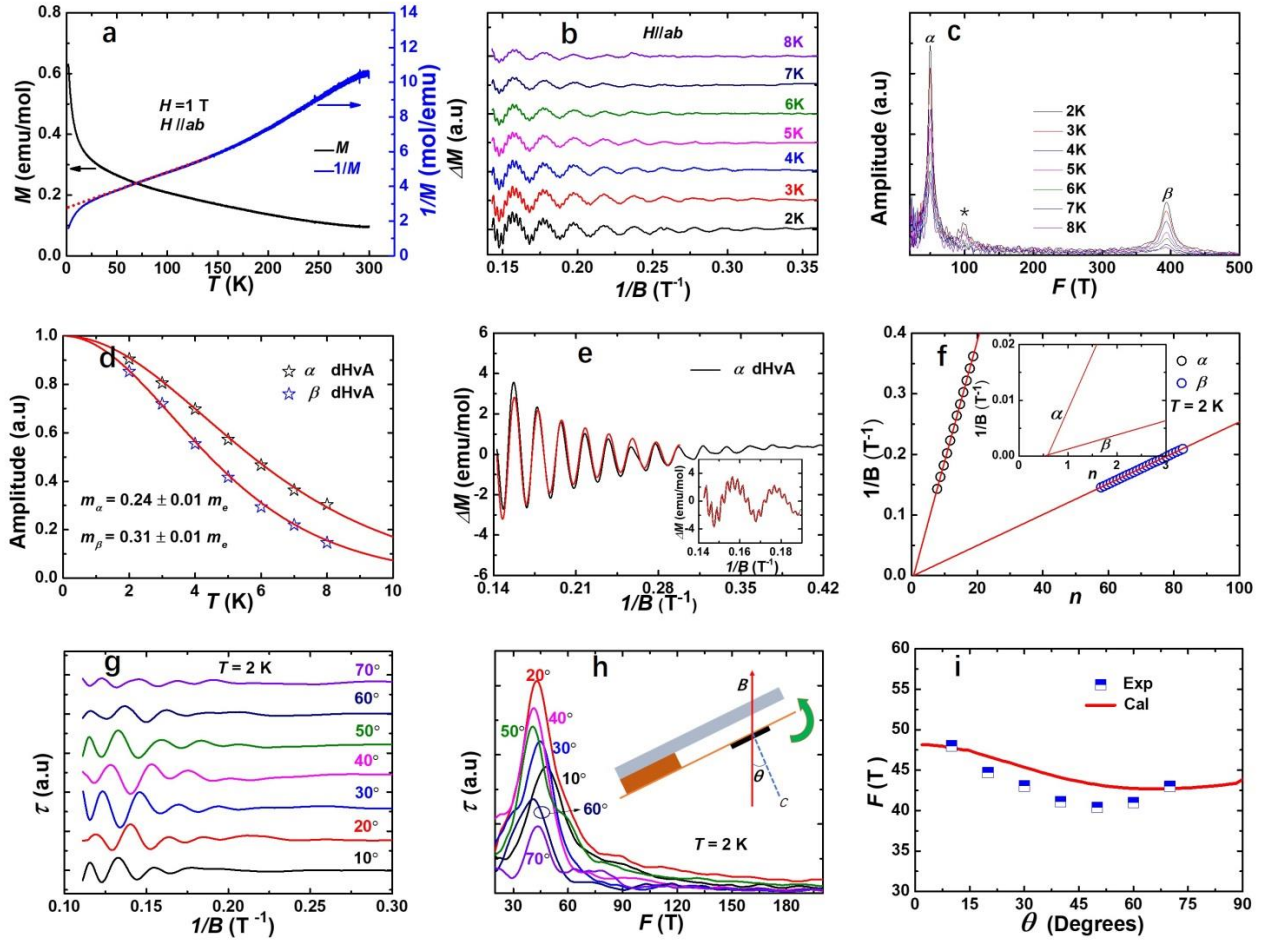

**Figure 3.** Quantum oscillations of NiTe$_2$. **a** The $T$-dependence of the magnetization (left) and the reciprocal magnetization (right) at 1 T field ($H \parallel ab$). The red dotted line is a fit to the Curie-Weiss law. **b** The magnetization oscillations at different temperatures after subtracting the background. **c** The corresponding FFT spectrum. A small peak marked by the asterisk only appears below 4 K. **d** The FFT amplitude as a function of temperature and the fits to $R_T$ to determine the effective masses. **e** The oscillations filtered from the $\alpha$ band (the black curve) and its fit (the red curve) to LK formula. The inset shows the fit to the raw data with the superposition of two LK oscillatory terms. **f** The Landau's fan diagram for the two major frequencies. The inset enlarges the intersection. **g** The torque oscillations at different angles at $T = 2$ K. **h** The FFT of the torque. The inset schematically shows the experimental setup. **i** The angular dependence of FFT frequency as determined from panel **h**. The red line is the theoretical calculation results (see main text).

in Eq. (1) is the Berry phase and $\delta$ is an additional phase shift determined by the dimensionality of Fermi surface. That is, $\delta=0$ ($\delta=\pm 1/8$) for 2D (3D) Fermi surfaces. From LK formula, the effective mass $m^*$ for each band can be obtained through the fit of the temperature dependence of its corresponding FFT amplitude to the thermal damping factor $R_T$. The effective masses for the $\alpha$ and $\beta$ bands are extracted to be 0.24 $m_e$ and 0.31 $m_e$ respectively, as shown in Figure 3d. One can further extract the Berry's phase for each individual band using two strategies. First, we may directly fit the oscillatory signals to the LK formula described above. Since the dHvA oscillations are only comprised of two major frequencies, it is possible to fit the data to the superposition of two LK oscillatory terms, i.e., one for each cyclotron frequency, as shown in the inset of

**TABLE I.** Parameters obtained from dHvA oscillations of a NiTe$_2$ single crystal. $F$ is the frequency of the oscillations. $m^*$ and $m_e$ are the effective electron mass and the bare electron mass, respectively. $T_D$ is the Dingle temperature. $\tau_q$ ($= \hbar/2\pi k_B T_D$) is the quantum relaxation time and $\mu$ is the quantum mobility.

| NiTe$_2$ | | dHvA ($H \parallel ab$) | | | | | $\phi_B$ | | |
|---|---|---|---|---|---|---|---|---|---|
| Methods | band | $F(T)$ | m*/m$_e$ | $T_D$ (K) | $\tau_q$ (ps) | $\mu$(cm$^2$/Vs) | $\delta$ = -1/8 | $\delta$ = 0 | $\delta$ = 1/8 |
| One-band fitting | $\alpha$ | 50 | 0.24 | 29 | 0.042 | 308 | 0.37$\pi$ | 0.62$\pi$ | 0.87$\pi$ |
| | $\beta$ | 394 | 0.33 | 15.1 | 0.082 | 437 | 1.1$\pi$ | 1.35$\pi$ | 1.6$\pi$ |
| Two-band fitting | $\alpha$ | 50 | 0.24 | 37 | 0.033 | 242 | 0.9$\pi$ | 1.18$\pi$ | 1.43$\pi$ |
| | $\beta$ | 394 | 0.33 | 15.6 | 0.033 | 421 | 1.08$\pi$ | 1.33$\pi$ | 1.58$\pi$ |
| Landau fan diagram | $\alpha$ | 49 | | | | | 0.95$\pi$ | 1.2$\pi$ | 1.45$\pi$ |
| | $\beta$ | 389 | | | | | 0.77$\pi$ | 1.02$\pi$ | 1.3$\pi$ |

Figure 3e. Alternatively, we may separate the oscillations associated with each band by applying the band filter in the FFT as the $\alpha$ and $\beta$-band frequencies are quite different. For example, the single band oscillations from the $\alpha$ band, represented by the black line in Figure 3e, were deconvoluted from the total oscillatory data using a FFT filter method. We further fit the $\alpha$ band oscillations to the LK formula, as seen from the red line in Figure 3e, to obtain its Berry's phase and the Dingle temperature $T_D$ (quantum relaxation time $\tau_q$). Second, the Berry's phase can also be extracted from the Landau's fan diagram as shown in Figure 3f. As known,[61] the valley in the magnetization should be assigned with a Landau level (LL) index of $n - 1/4$. The red lines in Figure 3f are the linear fits of the LL indices for $\alpha$ band and $\beta$ band. The slope of the linear fit is 49 T (389 T) for $\alpha$ band ($\beta$ band), in good agreement with the frequencies obtained from the FFT analysis. The inset of Figure 3f is an enlarged view near $1/B \to 0$. The intercept at $n$ abscissa from the linear fit is 0.59 (0.51) for $\alpha$ band ($\beta$ band), corresponding to a nontrivial Berry phase of $\phi_B = 2\pi|0.59+\delta|$ for $\alpha$ band ($\phi_B = 2\pi|0.51+\delta|$ for $\beta$ band). The fitting parameters from different methods are listed in Table I and *all* these methods consistently give the non-trivial Berry's phases for both $\alpha$ and $\beta$ bands.

To gain additional information on the 3D Fermi surface morphology, we implement the angle-dependent torque magnetometry as a probe which basically measures the magnetization of the sample as $\tau = M \times H$. Torque magnetometry measurements were carried out using a piezoresistive cantilever which was mounted on a rotation probe. The torque oscillations at $T$ = 2 K for various field-angles are illustrated in Figure 3g. The angle configurations are defined in the inset to Figure 3h where $\theta$=90° corresponds to $H \parallel ab$. It is noted that no torque oscillations were observed either in $\theta$=0° or $\theta$=90° configurations. This is not surprising because the magnetic torque is the first derivative of the free energy with respect to angle, and the free energy is extremal when field is along the high-symmetry directions (so the torque is zero at these angles in principle). The FFT analysis of the torque is given in Figure 3h. Only one peak centered around 45 T from the $\alpha$ band is detected in the magnetic torque whereas the $\beta$ band frequency is absent here. The angle variation of the frequency from the torque measurement is plotted in Figure 3i, together with the calculated frequencies from the first principles (see below). From Figure 3i, the frequency from the torque varies little with the angle, suggesting a rather 3D nature of the $\alpha$ band, consistent with the band calculations.

To identify the contributing Fermi pocket to each oscillation frequency, we calculate the electronic structure and the Fermi surface of NiTe$_2$. Figure 4a shows the bulk band structure of NiTe$_2$ along the $\Gamma$-$M$-$K$-$\Gamma$-$D$-$A$-$L$-$H$-$A$ direction. Two hole-dominant bands (band1 and band2) are plotted in red color while two electron-bands (band3 and band4) are shown in blue color. Band1 and band2 touch each other at the Dirac point $D$ in the $\Gamma$-$A$ direction. The Dirac energy is about 0.08 eV above the Fermi level. Along the in-plane (S-D-T) direction, Dirac cone is untilted (data not shown), while along the out-of-plane(A-$\Gamma$-A) direction the Dirac cone is titled. This suggests the characteristic feature of the type-II Dirac fermions as reported in PtTe$_2$ and PdTe$_2$. [20,22,23] By our calculations, the Dirac points are located at $k$ = (0, 0, ±0.35), slightly different from (0, 0, ±0.346) in PtTe$_2$ [22] and (0, 0, ±0.40) in PdTe$_2$ [23]. The Fermi surfaces (FSs) at *Dirac energy* in the bulk BZ is shown in Figure 4b where high-symmetry points and Dirac point (*D*) are indicated. In the full BZ, there are four bands crossing the Dirac energy level. It is shown that there is a spindle shaped hole pocket in the BZ center, while the much more complicated electron and hole pockets are composed of a series of large outer pockets. The hole pocket and electron pockets touch each other at two Dirac points along the $k_z$ direction. The calculated surface state spectrum along the $\overline{K}$-$\overline{\Gamma}$-$\overline{K}$ direction is shown in Figure 4c. Surface states around 0.5 eV are pointed by white arrows. A conical dispersion is located at a deep binding energy of 1.6 eV (pointed by green arrow). The cone-like dispersion corresponds to sharp surface states which connect the gapped bulk bands. We also carried out a surface spectral FS map as shown in Figure 4d and we can clearly see a hole-like pocket around BZ center, and extra surface states (bright lines pointed by white arrows) connect the bulk states between $\bar{\Gamma}$-$\bar{M}$ and $\bar{\Gamma}$-$\bar{K}$.

To figure out the contribution of each pocket to the dHvA oscillation mode, we calculate the extremal surfaces of each pocket with the field rotated from the *ab*-plane to the *c*-axis. The Fermi surface of band1 (Figure 4e) is a twisted hexagonal rod-like pocket and the DFT calculation gives a frequency range of 440 T to 2.0 kT and band masses from 0.2 to 0.8 $m_e$ when the field is rotated from $H \parallel ab$ to $H \parallel c$ (see Supplementary Information for the calculations of the angle-dependent dHvA frequency and band mass). As noted, the 440 T frequency should be responsible for the 400 T $\beta$ frequency seen in the dHvA measurements (Figure 3c). Consistently, this band is one of the two bands that form the type-II Dirac point shown in Figure 4a therefore possesses the non-trivial Berry's phase observed in the experiments. Band2 has a hexagonal barrel pocket and a small bowl-like electron pocket around A. The frequencies have a large range from 40 T to 21 kT and corresponding masses are from 0.8 to 3.5 $m_e$. We find that this small electron pocket around A, located at the center of the projected BZ for band2 in Figure 4e and of a frequency ~40 T, has the similar angle variation of the frequency to that of the nontrivial α mode, whose frequencies vary from 48 T to 40 T, as seen in Figure 3i. It is indicated that this pocket has nontrivial Dirac nature too. Band3 has a gear shaped electron pocket and its frequencies are from 1.8 kT to 21 kT with masses range from 0.8 to 6.1 $m_e$. For band4, which has six electron pockets at the BZ corners, the frequencies range is from 0.95 to 1.15 kT and band masses are from 0.27 to 0.32 $m_e$. The frequencies from band3 and band4 are not detectable in the dHvA experiments.

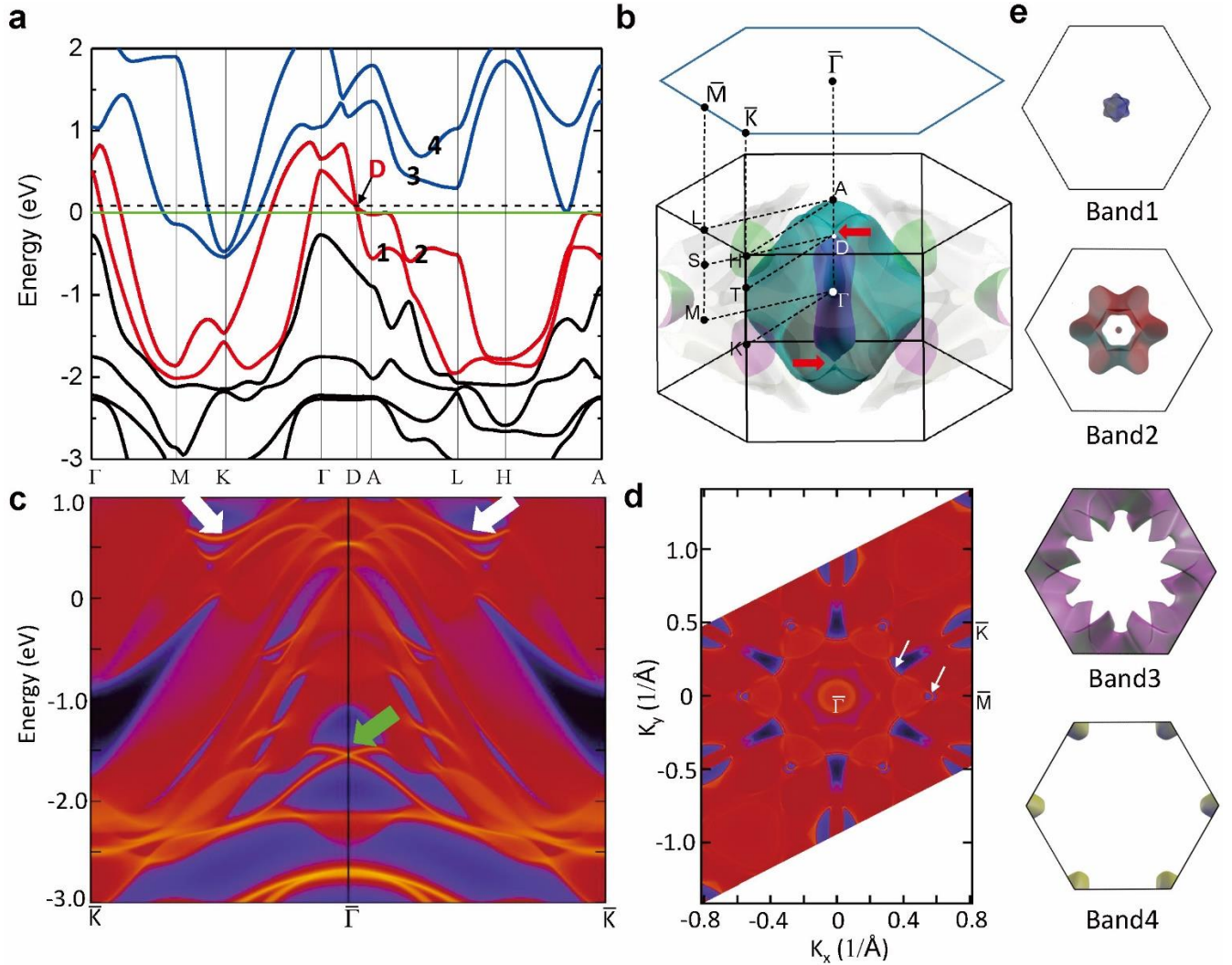

**Figure 4.** Band structure calculations of NiTe$_2$. **a** Calculated band structure of NiTe$_2$ with spin-orbit coupling along the Γ-*M*-*K*-Γ-*D*-*A*-*L*-*H*-*A* k-path. The green line guides for the Fermi level, while the dashed line guides for the Dirac point energy. "D" refers to the Dirac point. **b** Fermi surfaces at the Dirac point energy. The high symmetry points are marked in the BZ. Red arrows indicate the 3D Dirac points: D(0, 0, ±0.35). **c** Calculated surface state spectrum along the $\bar{K}$-$\bar{\Gamma}$-$\bar{K}$ directions using Wannier functions. **d** Calculated Fermi surfaces of (001) slab. **e** Individual Fermi surfaces of four bands crossing the Fermi level, projected onto the $k_x$-$k_y$ plane.

## 4. CONCLUSION

In summary, we have successfully synthesized the high quality single crystals of NiTe$_2$, a type-II Dirac semimetal in which a pair of Dirac points are located at ~80 meV above the Fermi level. Compared with its homologue PdTe$_2$ and PtTe$_2$, NiTe$_2$ may represent an improved platform to study the type-II Dirac physics as the Dirac fermions close to the Fermi energy imply more prominent contributions from these relativistic carriers in its transport and thermodynamic properties. Evidently, the quantum oscillations indeed reveal a nontrivial Berry's phase on the two major bands formed by Dirac fermions. More experiments, such as ARPES, need to identify the type-II Dirac nodes herein directly. In the future, it would be intriguing to explore the possible topological superconductivity in this type of materials by chemical substitution, intercalation or the high pressure. The intercalation of Ti or Cu may represent an effective route to this end.[62]

## ASSOCIATED CONTENT

### Supplementary Information

The XRD refinement results; Calculated oscillation frequencies and band masses for the four bands crossing the Fermi level as a function of field angle.

## AUTHOR INFORMATION

### Corresponding Author

* E-mail: xiaofeng_xu@cslg.edu.cn;   sankarndf@gmail.com

### Author Contributions

The manuscript was written through contributions of all authors. / All authors have given approval to the final version of the manuscript. / ‡These authors contributed equally.

### Notes

The authors declare no competing financial interest.

## ACKNOWLEDGMENT

The authors would like to thank Nigel Hussey, Zhiqiang Mao, Ali Bangura, Pabitra Biswas, C. M. J. Andrew for the fruitful discussion. This work is sponsored by the National Key Basic Research Program of China (Grant No. 2014CB648400), by National Natural Science Foundation of China (Grant No. 11474080, No. U1732162, No. 11504182, No. 11704047, No. 11374043). Xiaofeng Xu would also like to acknowledge the financial support from an open program from Wuhan National High Magnetic Field Center (2015KF15).